\documentclass[conference]{IEEEtran}

\usepackage{cite}
\usepackage{graphicx}
\usepackage{amsmath}
\usepackage{amssymb}
\usepackage{xcolor}
\usepackage{array}
\usepackage{booktabs}
\usepackage{tabularx}
\usepackage{multirow}
\usepackage{makecell}
\usepackage{comment}
\usepackage{url}
\usepackage{xurl}
\usepackage{pgfplots}
\usepackage{hyperref}
\pgfplotsset{compat=1.18}

\begin{document}



\title{Toward an Empirical Probabilistic Risk Manifestation Model of Organizational Cybersecurity in SMEs}

\author{
\IEEEauthorblockN{FNU Nurjahan,  Aidan Eiler, Mst Eshita Khatun, Lamine Noureddine, and Aisha Ali-Gombe}
\IEEEauthorblockA{
Division of Computer Science and Engineering\\
Louisiana State University\\
\{nurja1, aeiler3, mkhatu3, lnoureddine, aaligombe\}@lsu.edu
}
}

\IEEEoverridecommandlockouts
\makeatletter
\def\@IEEEpubidpullup{6.5\baselineskip}
\makeatother

\maketitle

\begin{abstract}
Small and medium-sized enterprises (SMEs) increasingly rely on digital infrastructure while operating under resource constraints that often result in inconsistent or underdeveloped cybersecurity practices. Although these organizations face growing cyber risk, it remains poorly understood how organizational security weaknesses empirically relate to downstream technical exposure, attack mechanisms, and outcomes. Prior work has largely examined organizational security conditions and technical security risk as separate problems, with limited empirical work quantifying the relationships between them. To address this gap, we present a cross-layer empirical study of organizational cybersecurity risk in SMEs, analyzing 281 validated security findings from 22 real-world SME cybersecurity assessments conducted over a two-year period through a pro-bono university cybersecurity clinic.
We first identify recurring organizational security functions through iterative thematic coding, then estimate an empirical Risk Manifestation Model linking these functions to exposure conditions, attack mechanisms, and cybersecurity outcomes, and use probability propagation to identify dominant risk pathways. The model characterizes empirical associations observed in this sample rather than causal or predictive relationships. Our analysis identifies eight organizational security functions associated with two exposure conditions, five attack mechanisms, and six outcome categories. Across most functions, the dominant pathway follows asset exposure to credential compromise to unauthorized access, whereas infrastructure and network security primarily propagates through network exposure; these pathways remain stable under leave-one-organization-out analysis. Finally, we evaluate whether SME cybersecurity assessments can be simplified while preserving meaningful security coverage. Retaining six functions reduces assessment burden by 24\% while preserving 97\% of critical findings and 92\% of risk-pathway coverage — a security-oriented reduction — while retaining five functions reduces burden by 45\% while preserving 89\% of critical findings and 85\% of risk-pathway coverage, a more efficiency-oriented alternative.
\end{abstract}


\IEEEpeerreviewmaketitle

\section{Introduction}
\label{section:introduction}
Small and medium-sized enterprises (SMEs) form a substantial part of the global economy, accounting for more than 90\% of businesses worldwide \cite{chidukwani2022}. Across OECD countries, SMEs represent approximately 99\% of firms and generate roughly 50--60\% of value added \cite{oecd_sme}; in the United States, small businesses account for 99.9\% of firms and employ more than 62 million people \cite{sba_small_business_faq}. Their economic importance is accompanied by growing cybersecurity exposure. Recent incidents illustrate the consequences: a 2024 ransomware attack against Young Consulting exposed sensitive insurance and medical information of more than 950,000 individuals \cite{alder2024_youngconsulting}, while the MediSecure breach exposed prescription and personal information of approximately 12.9 million individuals \cite{medisecure_incident_2024}. Such incidents are particularly concerning for SMEs, where limited resources and defensive capacity can make security weaknesses difficult to identify and remediate \cite{ponemon2019,clark2025}. Prior research has established that SME cybersecurity is shaped by resource constraints, limited expertise, informal processes, governance challenges, and difficulties in security decision-making \cite{chidukwani2022,rawindaran_2023,hoong2024navigating,wolf2021security}. Organizational cybersecurity studies further show that business priorities, staffing, governance, and deployment practices influence how security controls are implemented and maintained \cite{ruth2025perfect}. In parallel, cybersecurity measurement research quantifies technical conditions — vulnerability exposure, scanner/detection performance, malware prevalence, patching behavior — using operational or tool-based technical evidence \cite{kotzias2019mind, chalvatzis2019evaluation, tsiodra2023cyber}. These prior works largely treat organizational and technical security as separate problems: (1) research on SME security practices characterizes the organizational conditions under which controls succeed or fail, typically without connecting those conditions to downstream technical outcomes; (2) research on technical security risk measures exposure, attack surface, and outcome states, typically treating their organizational root causes as unmeasured or exogenous. What remains missing is not just a qualitative link between the two, but an empirically quantified one: \emph{how strongly, and through what pathways, do recurring organizational weaknesses in SMEs propagate into exposure conditions, attack mechanisms, and cybersecurity outcomes — and can that structure be measured with enough confidence to act on?}

This gap has practical weight for SMEs, who rarely have the resources to conduct exhaustive assessments and need a principled basis for prioritizing what to evaluate. We address it through a cross-layer empirical study of real-world SME cybersecurity assessments, building a four-layer Risk Manifestation Model (function → exposure → attack → outcome) and estimating the probabilistic dependencies that propagate risk across it. We then use this quantified structure to evaluate whether a reduced subset of assessed functions can preserve the findings, severe weaknesses, and risk-pathway coverage of a full assessment — offering SMEs an evidence-based basis for scoping cybersecurity assessments. We thus ask the following questions: {RQ1:} Which organizational security functions are consistently observed across cybersecurity assessments of SMEs? {RQ2:} What evidence-supported dependencies connect organizational security functions to exposure conditions, attack mechanisms, and cybersecurity outcomes, and how can these dependencies be quantified into a Risk Manifestation Model? {RQ3:} When propagated across the full model, what dominant risk-manifestation pathways emerge, and how robust are these pathways to variation in the underlying assessment data? {RQ4:} To what extent can a reduced subset of organizational security functions preserve the security coverage of a full SME assessment?

To answer these questions, we analyze 281 validated security findings documented in 22 real-world SME cybersecurity assessments conducted over a two-year period through a pro-bono university cybersecurity clinic. These assessments incorporated interviews, surveys, vulnerability scanning, configuration analysis, and penetration testing to gather technical and contextual evidence; our analysis is based on the resulting documented findings rather than on the primary interview or survey data. We first apply iterative thematic coding to identify recurring organizational security functions. We then derive evidence-supported relationships among organizational functions, exposure conditions, attack mechanisms, and outcomes and aggregate them into a Risk Manifestation Model. Probability propagation over this model identifies function-specific downstream distributions and dominant risk-manifestation pathways, whose robustness and sampling uncertainty are evaluated through organization-level resampling. Finally, we evaluate progressively reduced assessment subsets using retained findings, critical findings, and probabilistic risk-pathway coverage under alternative function prioritization criteria. The modeled relationships represent empirical associations rather than causal effects, and the model is intended to characterize observed risk-manifestation structure rather than to predict outcomes for organizations outside the studied sample. 

Our analysis identifies eight organizational security functions associated with two exposure conditions, five substantive attack mechanisms, and six cybersecurity outcome categories. Seven functions share the dominant pathway \emph{Asset Exposure $\rightarrow$ Credential Compromise $\rightarrow$ Unauthorized Access}, whereas Infrastructure \& Network Security primarily propagates through Network Exposure. The dominant complete pathways remain unchanged across all 176 leave-one-organization-out (LOO) function--exclusion comparisons. The assessment reduction analysis further identifies security-oriented and efficiency-oriented operating points, which remain consistent under alternative function prioritization criteria. 

In summary, this paper makes the following contributions:
\begin{itemize}
\item \textbf{Empirical Characterization of SME Organizational Security.}
We derive eight recurring organizational security functions from validated findings across 22 real-world SME assessments and characterize their prevalence and severity.
\item \textbf{Empirical Risk Manifestation Model.}
We derive an empirical probabilistic model from evidence-supported assessment relationships linking organizational security functions, exposure conditions, attack mechanisms, and cybersecurity outcomes.
\item \textbf{Probabilistic Risk Propagation and Robustness.}
We characterize function-specific downstream risk distributions and dominant complete pathways and evaluate their robustness and sampling uncertainty through organization-level resampling.
\item \textbf{Evidence-Based Assessment Reduction.}
We quantify the trade-off between assessment-question burden and retained security coverage and compare function-reduction strategies based on overall and critical-finding prevalence.
\end{itemize}

\section{Background} 
\label{section:background}

\par\noindent {\bf Key Definitions.} To ensure conceptual clarity, we define key terms used in this study.
\par\noindent \textit{(I) Organizational Security Function} refers to a recurring domain of organizational security practice — such as access control, asset management, or vulnerability management — in which one or more security weaknesses were identified during an SME assessment. In this study, organizational security functions are derived empirically through thematic coding of assessment findings (Section~\ref{section:method}) and represent domains of observed organizational deficiency rather than a general-purpose taxonomy of security capabilities.
\par\noindent \textit{(II) Exposure Conditions} refers to organizational or system conditions—such as configuration, deployment context, or operational practices—that make a vulnerability or weakness accessible to potential exploitation.
\par\noindent\textit{(III) Attack Mechanism} refers to the technical method or vector by which an exposure condition is exploited, e.g., credential theft, phishing, or lateral movement.
\par\noindent \textit{(IV) Cybersecurity outcomes} refer to the observable technical and organizational consequences of a security weakness being exploited or manifesting, including unauthorized access, data compromise, operational disruption, financial loss, reputational harm, and regulatory or legal impact.
\par\noindent\textit{(V) Impact} captures the potential consequences of exploitation on confidentiality, integrity, availability, and organizational operations.
\par\noindent\textit{(VI) Risk} reflects the combination of the likelihood of exploitation and the resulting impact; in this study, likelihood is captured through empirically estimated conditional dependencies between layers, while impact is reflected in outcome severity and finding criticality.
 
\par\noindent\textbf{Cybersecurity Assessment in Resource-Constrained SMEs.} SMEs often operate with limited financial resources, cybersecurity expertise, and dedicated security staff, leaving security responsibilities to general IT personnel or business owners \cite{rawindaran_2023, curtin_2025, wong2022role, enisa_sme, chidukwani2022}. Consequently, foundational practices such as asset management, access control, secure configuration, vulnerability management, monitoring, and business continuity may be implemented inconsistently. Increasing reliance on cloud services, third-party platforms, remote access, and internet-connected devices further expands attack surfaces and management complexity \cite{chidukwani2022}. Frameworks such as the NIST Cybersecurity Framework provide structured guidance across Identify, Protect, Detect, Respond, and Recover \cite{nist_sme}, but such frameworks are largely prescriptive and do not empirically link organizational practice gaps to the technical exposure, attack, and outcome conditions they are intended to prevent — nor do they indicate which practices matter most when full implementation is infeasible. These conditions motivate examining how recurring organizational weaknesses observed during SME assessments correspond to downstream technical security conditions and risk.

\section{Related Work}
\label{section:relatedwork}

\par\noindent {\bf Studies about Organizational Cybersecurity Assessment in SMEs.}
SMEs face cybersecurity challenges shaped by limited financial resources, shortages of security expertise, governance constraints, and competing business priorities \cite{chidukwani2022, rawindaran_2023, hoong2024navigating, wolf2021security, wilson2023won}. These constraints affect core organizational security functions, including access control, asset and vulnerability management, monitoring, security awareness, and business continuity. To improve cybersecurity readiness in such environments, prior work has proposed SME-focused policies, simplified assessment frameworks, maturity models, and practical guidance \cite{enisa_sme_page, curtin_2025, pawar_2022, rawindaran_2025, manzoor2024cybersecurity}. Large-scale surveys nevertheless show persistent gaps in foundational practices such as security documentation, incident response, and business continuity planning \cite{uk_breach_survey_2024}. Other studies emphasize that cybersecurity adoption also depends on management commitment, organizational flexibility, vendor relationships, cyber hygiene, proactive engagement, and access to external expertise \cite{khan_2025, hasani2023evaluating, huaman2021large, sadok_2020}. Systematic reviews similarly find that SME cybersecurity research has focused primarily on awareness, governance, and resource constraints rather than empirically linking organizational practices to measurable technical security outcomes \cite{junior2025unawareunfundeduneducatedsystematic}. Overall, prior work establishes the importance of organizational capability in SME cybersecurity, but provides limited empirical evidence on how recurring organizational security weaknesses correspond to downstream exposure conditions, attack mechanisms, and cybersecurity outcomes. Our work addresses this gap using real-world cybersecurity assessment findings to derive an empirical cross-layer characterization of these relationships.

\par\noindent\textbf{Probabilistic Modeling of Cyber Risk.}
Probabilistic models have been widely used to characterize cyber risk by representing dependencies among vulnerabilities, attack paths, and network compromise. Bayesian networks and attack graphs have been combined to estimate attack likelihoods, support cyber risk assessment, and prioritize defensive actions \cite{frigault_2008, Poolsappasit_2012, munoz_2016}. More recent approaches incorporate network topology, known vulnerabilities, and security alerts to dynamically update compromise probabilities during ongoing attacks \cite{francois-xavier_2016}. Collectively, these approaches provide effective mechanisms for modeling technical cyber risk, but they typically rely on expert-defined dependency structures, attack graphs, or network topology as their primary inputs. In contrast, our approach derives both the dependency structure and the associated conditional probabilities empirically from real-world cybersecurity assessment findings.

\par\noindent{\bf Connecting Organizational Security with Technical Risk.}
Prior research shows that technical security outcomes are shaped not only by technical controls but also by the organizational practices governing their implementation and maintenance \cite{gale2022governing, hoong2024navigating, alraja2023information, uchendu2021developing}. Real-world incident studies similarly indicate that compromises often arise from operational weaknesses rather than sophisticated attack techniques \cite{dietrich_2018_investigating}. This relationship appears across multiple domains: vulnerability management is affected by usability, awareness, and automation limitations \cite{ayala_2025_mixed_method}; cloud security by documentation, tool usability, and operator knowledge gaps \cite{hashmi2026mapping}; asset visibility by incomplete inventories that leave systems untracked and unpatched \cite{ethembabaoglu2024unpatchables, kustosch2025patching}; patch management by organizational processes that prolong exposure \cite{dissanayake2021grounded, dissanayake_patch}; and authentication by persistent weaknesses that increase credential-related risk \cite{gavazzi2023study, nisenoff2023two, thomas2019protecting, klivan_2023}. Collectively, these studies show that control effectiveness depends on how security is implemented, maintained, staffed, coordinated, and governed \cite{woods_seymour, dissanayake_patch, ethembabaoglu2024unpatchables}. However, this evidence remains distributed across single-domain studies — such as vulnerability management, cloud security, patch management, and authentication — rather than integrated into a unified, quantified structure spanning organizational functions through to technical outcomes. We address this by deriving the Risk Manifestation Model — an empirical probabilistic model estimated from real-world assessment findings — that quantifies conditional relationships across organizational security functions, exposure conditions, attack mechanisms, and cybersecurity outcomes. 
\section{Method}
\label{section:method}
This study adopts a cross-layer empirical approach to examine how organizational security weaknesses are associated with exposure conditions, attack mechanisms, and cybersecurity outcomes in resource-constrained SMEs, typically small, non-IT-centric organizations without dedicated security staff. The methodology integrates qualitative thematic analysis of security assessment findings with probabilistic analysis of empirically observed relationships among organizational security functions, exposure conditions, attack mechanisms, and cybersecurity outcomes, providing a structured framework for characterizing how organizational security weaknesses correspond to downstream cybersecurity conditions and outcomes. This study is based on secondary analysis of documented security assessment findings. While the underlying assessments incorporated contextual data collection (e.g., interviews, surveys) alongside technical evaluation methods, the analysis in this study was conducted on the resulting validated findings rather than on raw contextual inputs.

\subsection{Study Context: Cybersecurity Clinic Model} The XXX Cybersecurity Clinic \emph{(name omitted for double-blind review)} is an experiential learning program designed to provide students with hands-on cybersecurity training through community engagement. Established in Fall 2023 through grant funding, the clinic delivers pro-bono cybersecurity services to SMEs, particularly those with limited internal security expertise and limited budget for third-party security services. Services include comprehensive security assessments, educational seminars, and one-on-one cybersecurity counseling. The primary clients of the clinic are small, non-IT-centric businesses. Organizations may engage with the clinic directly or be referred through external partners such as the Small Business Development Center (SBDC) \cite{sbdc}. These engagements provide a real-world setting in which students perform structured security evaluations under faculty supervision, producing documented assessment reports that serve as the basis for this study's analysis.

\subsection{Data Source and Scope} This study is based on a retrospective analysis of cybersecurity assessment reports generated by the clinic over a two-year period. The dataset consists of reports from 22 SMEs. Each report contains validated security assessment findings derived from a combination of technical evaluation methods — including vulnerability scanning, configuration analysis, and penetration testing — and contextual evidence gathered through interviews and surveys, assessed against established control frameworks such as the CIS Controls. These assessment findings constitute the core unit of analysis in this study.

\subsection{Assessment and Data Generation Pipeline}
The clinic adopts the Center for Internet Security (CIS) Critical Security Controls (CIS Controls v8) \cite{ciscontrols} as the primary assessment framework, encompassing 18 controls. Table~\ref{tab:org_characteristics} summarizes the characteristics of the 22 organizations included in our dataset. To preserve confidentiality, all organizations are anonymized and referred to using identifiers (O1--O22). The organizations span multiple sectors, including construction, healthcare, consulting, retail, manufacturing, education, and government contracting. Organizational sizes range from single-person businesses to approximately 200 employees, all of which fall within the definition of SME \cite{eu_sme_definition}. 

\begin{table}[t]
\centering
\small
\caption{Characteristics of participating organizations (anonymized).}
\label{tab:org_characteristics}
\setlength{\tabcolsep}{3pt}
\begin{tabularx}{\columnwidth}{
p{0.5cm}
>{\raggedright\arraybackslash}X
p{1.4cm}
p{1.5cm}
p{1cm}}
\toprule
\textbf{Org} & \textbf{Sector} & \textbf{Employees} & \textbf{Framework} & \textbf{Est.} \\
\midrule
O1  & Marine Construction        & 11--50   & CIS            & 2005 \\
O2  & Finance / Consulting       & 2--10    & CIS            & 2015 \\
O3  & Construction               & 10--20       & CIS            & 2008 \\
O4  & Healthcare                 & 1--10       & CIS            & 2018 \\
O5  & Equipment Retail           & 15--25       & CIS            & 1982 \\
O6  & Architecture / Design      & 11--50   & CIS            & 1949 \\
O7  & Landscaping                & 160--200 & CIS            & 2001 \\
O8  & Inspection / Construction Services & 1--10 & CIS & 2018 \\
O9  & Dry Cleaning and Laundry Service & 25--200 & CIS & 1900 \\
O10 & Roofing / Construction     & 11--50   & CIS            & 1978 \\
O11 & Manufacturing              & 1--10    & CIS            & 1997 \\
O12 & Food Distribution          & 51--200  & CIS            & 1978 \\
O13 & Professional Services      & 11--200  & CIS            & 1953 \\
O14 & Engineering                & 51--200  & CIS            & 2001 \\
O15 & Healthcare                 & 1--10   & CIS            & 2024 \\
O16 & Marketing / Creative       & 1--20    & CIS            & 2006 \\
O17 & Education / Childcare      & 11--50   & CIS            & 2015 \\
O18 & Healthcare / Wellness      & 1--10    & CIS            & 1985 \\
O19 & Healthcare                 & 3--10    & CIS            & 2016 \\
O20 & Defense Services           & 1--10    & CIS → CMMC L1  & 2003 \\
O21 & Consulting                 & 20--30       & CIS → CMMC L1  & 2005 \\
O22 & Defense Services           & 10--20   & CIS → CMMC L1  & 2016 \\
\bottomrule
\end{tabularx}

\end{table}

Among the 22 participating organizations, three required alignment with Cybersecurity Maturity Model Certification (CMMC) Level 1 certification requirements \cite{cmmc_level1_2024}, and their assessments incorporated the corresponding CMMC security practices. Organizations requesting a comprehensive assessment complete an intake process followed by the establishment of a Memorandum of Understanding (MoU) and Rules of Engagement (RoE). Assessments are conducted over a period of approximately 4--7 weeks. The assessment begins with a kickoff interview, conducted either in person or remotely via Zoom or Microsoft Teams, that gathers responses to an initial subset of the CIS Controls v8 assessment questionnaire — covering the organization's IT infrastructure, network architecture, governance practices, and compliance requirements. Because the complete instrument comprises 93 standardized questions spanning all 18 CIS Controls, the remaining questions are collected through a follow-up survey/questionnaire sent after the kickoff interview. Together, the kickoff interview and follow-up questionnaire are designed to elicit responses to the full standardized 93-question instrument for every client, ensuring consistency across assessments. Like the kickoff interview, individual client responses to the questionnaire are used within clinic operations to guide the technical assessment but are not directly analyzed in this study; however, the questionnaire's structure — specifically, its mapping of each of the 93 standardized questions to the corresponding CIS Control, which follows the publicly available CIS Controls v8 documentation \cite{ciscontrols} — is used in Section~\ref{method:m4} to relate validated findings to assessment burden. 

Technical assessments are performed using a combination of tools installed on a secure jump server and connected to the client environment. This phase includes network and vulnerability scanning (e.g., Nmap \cite{nmap}), configuration analysis using CIS-CAT Pro\cite{ciscat} for benchmarking, and short-term event monitoring using Wazuh \cite{wazuh}. For one organization whose infrastructure was primarily hosted on Amazon Web Services, an additional cloud security assessment was conducted using Prowler \cite{prowler}, an open-source tool that evaluates cloud security configurations against industry best practices and provides remediation guidance for identified vulnerabilities. These tools generate technical artifacts such as vulnerability findings, configuration gaps, and indicators of exposure, which are validated and consolidated into assessment reports. 

Assessments were conducted by structured undergraduate teams comprising red, green, and blue team roles under faculty supervision. All findings were reviewed and validated by faculty advisors with academic and industry cybersecurity expertise prior to inclusion in final reports. Across the collected reports, a total of 281 security assessment findings were identified and consolidated for analysis. Findings were recorded on a per-client basis; while similar finding types may recur across organizations, each record represents a distinct client-specific observation. Assessment scope was constrained by client permissions. For example, phishing, social engineering, and denial-of-service (DoS) testing were excluded in some engagements. The final reports used as the data source include detailed descriptions of identified vulnerabilities, affected assets, mappings to CIS Controls, associated risk ratings, and recommended remediation actions.


\subsection{Data Processing and Anonymization}
Prior to analysis, all assessment reports were de-identified to remove organization names, contact details, and other potentially identifying information, and each organization was assigned a unique identifier. The reports were then systematically processed to extract relevant technical findings, including vulnerability data, severity scores, and control-level gaps. Only de-identified data were used for analysis; while findings are tracked at the organization level to support per-organization analyses, no organization's real-world identity is disclosed or inferable from the reported results.

For this study, data were primarily extracted from the findings sections of each of the 22 assessment reports. A \textit{finding} is a security-relevant observation identified during the cybersecurity assessment that indicates the presence of a vulnerability, misconfiguration, or control weakness, supported by technical or procedural evidence. Each identified finding was documented using a structured template capturing technical weaknesses, contextual conditions, risk attributes (risk rating, likelihood and impact), and supporting metadata such as remediation guidance and evidence sources as illustrated in Table~\ref{tab:finding_template}. 

Within this structure, each finding is assigned a unique identifier, a concise title, and a description field that provides detailed information about the observed security weakness. This may include identified CVEs or other security conditions, such as lack of MFA, contextualized for the specific client organization being assessed. The risk profile on the other hand provides a risk rating for the identified security weakness based on the likelihood of exploitation given the organization's defensive context and the potential impact of such exploitation on the client organization. The structure also includes remediation steps, estimated effort, and additional contextual information, such as affected assets and mappings to the CIS Controls.

\begin{table}[t]

\caption{Simplified structured representation of security findings; a complete example is provided in Appendix~\ref{sec:illustrative_examples}.}
\label{tab:finding_template}
\centering
\begin{tabular}{p{3cm} p{5cm}}
\toprule
\textbf{Category} & \textbf{Fields and Example Values} \\
\midrule

\textbf{Identifier} 
& ID F002 \\

\textbf{Finding} 
& No MFA enabled \\

\textbf{Finding Description} & No MFA enabled on Google Drive and QuickBooks accounts; frequent phishing exposure \\
\hline

\textbf{Risk Attributes} 
& \textit{Risk Rating}: High \\

\textbf{Likelihood} 
& \textit{Likelihood Rating}: Likely; \newline{\textit{Likelihood Description}: Phishing likelihood increases due to absence of MFA} ;  \\
\textbf{Impact} 
& \textit{Impact Rating}: High; \newline \textit{Impact Description}: Potential client data compromise and financial loss. \\\hline

\textbf{Mitigation} 
& \textit{Remediation Effort}: Low; 
\textit{Remediation}: Enable MFA on business-critical platforms \\\hline

\textbf{Context} 
& \textit{Affected Assets}: Google Drive; QuickBooks \newline
\textit{Control}: CIS Control 6.3; \textit{Evidence}: Client interview; \textit{Reference}: NIST MFA guidance \\

\bottomrule

\end{tabular}

\end{table}


\subsection{Analysis Methodology}

\subsubsection{Qualitative Analysis}\label{method:m1} To characterize organizational security weaknesses in our dataset, we applied thematic coding to the \emph{Finding Description} field. Thematic analysis is a qualitative approach commonly used to identify recurring patterns and themes in textual data\cite{braun2006thematic}. An iterative thematic coding process \cite{saldana2021coding} was employed to identify recurring organizational security weaknesses across the extracted security assessment findings. Although the underlying security assessments were conducted using established cybersecurity controls, the objective of this study was not to classify findings according to the assessment framework. Instead, we sought to identify recurring organizational weakness patterns that emerged across organizations and were suitable for subsequent probabilistic modeling. Consequently, the analysis adopted an inductive thematic approach, while established cybersecurity frameworks, including the CIS Controls, served as a \emph{sensitizing framework} during interpretation rather than as predefined coding categories. In the first coding cycle, open coding was applied to capture descriptive security issues directly from the assessment findings without imposing predefined categories. This stage focused on identifying observable technical and operational issues such as authentication weaknesses, missing security policies, outdated software, monitoring gaps, and physical or network exposure. These open codes provided a granular representation of the vulnerabilities, misconfigurations, and organizational deficiencies observed during the assessments. To provide transparency into the synthesis of the qualitative findings, Table~\ref{tab:theme_summary} presents the mapping between representative coding categories and the resulting organizational security functions. A detailed mapping of the initial coding labels to the final organizational security functions is provided in Appendix~\ref{sec:illustrative_examples} (Table~\ref{tab:theme_mapping}).

\begin{table}[t]
\centering
\small
\caption{Representative coding categories mapped to synthesized organizational security themes. The complete mapping of initial coding labels is provided in Appendix~\ref{sec:illustrative_examples}.}
\label{tab:theme_summary}
\begin{tabularx}{\columnwidth}{p{0.63\columnwidth}p{0.30\columnwidth}}
\toprule
\textbf{Representative Coding Categories} & \textbf{Synthesized Theme} \\
\midrule
Identity verification, credential management, authorization, account governance & Access Control \& Authentication \\

Software vulnerabilities, patch management, vulnerability remediation, security protocols & Vulnerability Management \\

Secure configurations, endpoint hardening, application security, removable media controls & Configuration Management \\

Data governance, encryption, secure storage, backup and recovery & Data Protection \& Resilience \\

Security policies, awareness training, incident response, third-party governance & Governance and Organizational Capacity \\

Network segmentation, boundary protection, secure protocols, external connectivity & Infrastructure \& Network Security \\

Audit logging, log management, monitoring, incident detection & Monitoring \& Detection \\

Hardware/software inventories, asset ownership, lifecycle management & Asset Management \\
\bottomrule
\end{tabularx}
\end{table}

In the second coding cycle, axial coding was used to group related open codes into higher-level pattern codes representing broader organizational security practices. During this stage, the emerging codes were iteratively compared and refined, with established cybersecurity control frameworks providing a sensitizing lens for interpreting relationships among related security issues. Rather than directly adopting predefined control families, the resulting pattern codes were derived from recurring empirical patterns observed across the assessment findings. This process resulted in twelve pattern codes: (1) Inventory of Assets and Software, (2) Data Resiliency, (3) Security Configuration, (4) Identity \& Access Management (IAM), (5) Continuous Vulnerability Management, (6) Activity Logging and Monitoring, (7) Endpoint Defenses, (8) Network Defenses, (9) Security Training, (10) Vendor Management, (11) Application Software Security, and (12) Incident Response Management. For example, findings describing missing asset inventories, unmanaged devices, or lack of software tracking were grouped under \textit{Inventory of Assets and Software}. Findings involving weak password policies, missing multi-factor authentication, and shared credentials were grouped under \textit{Identity and Access Management}. Similarly, findings related to outdated software, unpatched systems, and exposed services were grouped under \textit{Continuous Vulnerability Management}.

To improve coding consistency, a structured codebook was developed and iteratively refined throughout the coding process. The codebook defined each pattern code, specified inclusion criteria, and documented representative examples from the dataset. The primary researcher applied the codebook to the full dataset. To assess intercoder reliability, a second coder independently analyzed a randomly selected subset of the findings (20\% of the dataset) during both cycle 1 and cycle 2 pattern coding. Intercoder agreement was evaluated using Cohen's Kappa \cite{cohen1960coefficient} for the second-cycle pattern codes. The resulting Cohen’s Kappa value of $\kappa = 0.89$ indicates near-perfect agreement, demonstrating strong consistency in the application of the coding scheme. Discrepancies between coders were reviewed and resolved through discussion, and minor refinements were made to the code definitions to improve clarity and ensure consistent application across the dataset. The pattern codes were subsequently synthesized into eight organizational security functions representing systemic security weaknesses across organizations. This synthesis was reviewed by both coders, and consensus was reached on the final structure: (1) Access Control \& Authentication, (2) Configuration Management, (3) Vulnerability Management, (4) Governance and Organizational Capacity, (5) Infrastructure \& Network Security, (6) Data Protection \& Resilience, (7) Monitoring \& Detection, and (8) Asset Management. Because this final synthesis was achieved through discussion-based consensus between both coders rather than independent parallel coding, formal intercoder reliability was not computed for this stage; the high agreement already established at the pattern-code level ($\kappa = 0.89$) provided the basis from which consensus-based synthesis proceeded. These organizational security functions reflect recurring domains of security practice in which weaknesses were observed to emerge and persist across the assessed organizations.


\subsubsection{Risk Manifestation Model}\label{method:m2}
The eight organizational security functions identified through thematic coding characterize what recurring weaknesses were observed across the assessed SMEs, but not how those weaknesses relate to downstream technical risk. To capture this, we derive exposure conditions, attack mechanisms, and outcome categories from each finding's documented risk attributes — including their risk rating and contextual likelihood/impact descriptions — systematically analyzed for recurring patterns across the dataset. Together, these layers form the Risk Manifestation Model: an empirical, four-layer probabilistic structure that traces how organizational security functions correspond to exposure conditions, attack mechanisms, and cybersecurity outcomes. Rather than assuming these relationships a priori, the model's structure and conditional probabilities are both derived directly from evidence-supported associations documented in the assessment findings. We formalize this model as a probabilistic dependency graph, described next.

\par\noindent{\bf Probabilistic Dependency Modeling: }
The Risk Manifestation Model is formalized as a probabilistic dependency graph, $
G=(V,\mathcal{L}),
$ where $V$ represents the set of layer-specific states and $\mathcal{L}$ denotes weighted directed edges between adjacent layers. The vertex set is organized into four layers, $
V=F\cup E\cup A\cup O,
$ where $F$, $E$, $A$, and $O$ denote organizational security functions, exposure conditions, attack mechanisms, and cybersecurity outcomes, respectively. Edges are defined only between adjacent layers, $
\mathcal{L}\subseteq (F\times E)\cup(E\times A)\cup(A\times O),
$ forming a layered dependency structure from organizational security functions to downstream cybersecurity outcomes.

The exposure conditions, attack mechanisms, and cybersecurity outcomes were coded from the detailed finding descriptions and supporting contextual evidence documented in the assessment reports. Each finding was assigned to at least one exposure condition based on the technical and organizational context of the identified weakness. These annotations are non-exclusive: a single finding may support multiple exposure conditions, attack mechanisms, or cybersecurity outcomes. Importantly, adjacent-layer associations were derived from the technical and contextual evidence available for each finding. We therefore did not assume that every exposure condition assigned to a finding was associated with every attack mechanism, or that every attack mechanism was associated with every identified outcome. The resulting Function--Exposure, Exposure--Attack, and Attack--Outcome links represent analytical associations grounded in the assessment evidence rather than explicitly observed attack sequences. The dependency graph aggregates these finding-level associations across the dataset. Each edge weight is defined as $
w(u,v)=P(v\mid u),$ where $P(v\mid u)$ denotes the row-normalized empirical probability of a coded downstream association from parent state $u$ to child state $v$.
Accordingly, the conditional association distributions for $P(E\mid F)$,
$P(A\mid E)$, and $P(O\mid A)$ were estimated as:
\begin{equation}
P(Y=y_j\mid X=x_i)
=
\frac{
N_{\mathrm{assoc}}(x_i,y_j)
}{
\sum_k N_{\mathrm{assoc}}(x_i,y_k)
},
\label{eq:association_probability}
\end{equation}
where $N_{\mathrm{assoc}}(x_i,y_j)$ denotes the number of
evidence-supported associations between parent state $x_i$ and child state $y_j$, and the denominator is the total number of downstream associations originating from $x_i$. Because an individual finding may support multiple relationships, the number of associations may exceed the number of unique findings. The resulting probabilities therefore characterize the empirical distribution of supported downstream associations conditional on each parent state, rather than the proportion of findings assigned to a mutually exclusive child category. This estimation procedure is applied separately to the Function--Exposure, Exposure--Attack, and Attack--Outcome associations. The resulting conditional association distributions parameterize the dependency graph and provide the basis for the probability propagation analysis presented in RQ3.

\subsubsection{Probabilistic Risk Propagation} \label{method:m3}
The conditional association distributions estimated in Section \ref{method:m2} were combined to propagate probability mass through the layered dependency graph. Under this model, propagation characterizes how the empirical downstream association structure changes across organizational security functions. Consistent with the layered structure of the graph, we assume that each layer's downstream association distribution depends only on the immediately preceding layer (a first-order Markov-style assumption across layers): for instance, the attack-mechanism distribution is represented by $P(A\mid E)$ rather than by a separate distribution conditioned on both $E and F$. This assumption provides a parsimonious representation of the empirical relationships given the available sample size and allows probability mass to be propagated across non-adjacent layers by marginalizing over the intermediate layer. The propagated probability of an attack mechanism given an organizational security function is:
\begin{equation}
P(A\mid F)
=
\sum_{E}
P(A\mid E)\,
P(E\mid F).
\label{eq:attack_inference}
\end{equation}
Similarly, the propagated probability of a cybersecurity outcome given an organizational security function is:
\begin{equation}
P(O\mid F)
=
\sum_{E}\sum_{A}
P(O\mid A)\,
P(A\mid E)\,
P(E\mid F).
\label{eq:outcome_inference}
\end{equation}
To characterize complete risk-manifestation pathways, we computed the model probability of each specific exposure–attack–outcome combination for every organizational security function. Unlike the marginal quantities in Equations~\ref{eq:attack_inference}–\ref{eq:outcome_inference}, this joint pathway probability is evaluated for each individual $(E,A,O)$ combination rather than summed over intermediate states, allowing identification of the single most probable complete pathway for each function:
\begin{equation}
P(E,A,O\mid F)
=
P(E\mid F)\,
P(A\mid E)\,
P(O\mid A).
\label{eq:path_probability}
\end{equation}
For each organizational security function, the
exposure--attack--outcome combination with the highest model probability was identified as its dominant risk-manifestation pathway, i.e., we evaluated all feasible ($E,A,O$) combinations (up to $\mid E\mid\times\mid A\mid\times\mid O\mid$) and selected the maximum-probability combination as the dominant pathway. These dominant pathways form the basis of the comparative analysis presented in RQ3. The layered model assumes that downstream association distributions depend on the immediately preceding layer. For example, after conditioning on an exposure condition, the attack-mechanism distribution is represented by $P(A\mid E)$ rather than by a separate distribution for every organizational security function. This assumption provides a parsimonious representation of the empirical relationships given the available sample size. The propagated quantities should therefore be interpreted as relative pathway probabilities under the empirical association-transition model. They do not imply that every instance of a particular exposure condition leads to the same attack mechanism, nor do they estimate the probability that a future cyber incident will follow a particular sequence. Rather, they summarize recurring relationships supported by the assessment findings. More generally, the model captures empirical associations rather than causal relationships.

\par\noindent{\bf Robustness Analysis.}
We evaluated organization-level robustness using leave-one-organization-out (LOO) resampling. Each of the 22 organizations was independently excluded, and the complete probabilistic model was re-estimated using the remaining 21 organizations. For each organizational security function, we measured stability as exact agreement between the dominant attack mechanism, outcome, and complete pathway identified under the full-data model and those identified under each LOO re-estimation. Because each organization contributed multiple assessment findings, we treated the organization, rather than individual findings, as the resampling unit. Across the eight organizational security functions and 22 LOO iterations, this yielded 176 function-level robustness comparisons in total.

\par\noindent{\bf Uncertainty Estimation.}
The 22 LOO estimates of each full-data dominant-pathway probability were additionally used to compute jackknife standard errors and approximate 95\% confidence intervals \cite{jackknife}. This analysis quantifies sampling uncertainty in the estimated pathway probabilities while separately evaluating the stability of the dominant pathway structure.

\subsubsection{Assessment Reduction}\label{method:m4}
The preceding phases establish the organizational security functions, the Risk Manifestation Model linking them to downstream risk, and the robustness of the resulting pathways. This quantified structure enables a further question: if assessment resources are constrained, which functions can be prioritized — and which omitted — with the least loss of security coverage? We address this by constructing nested assessment subsets at the organizational-security-function level, using two prioritization strategies: a \emph{prevalence-based} strategy, which ranks functions by their number of validated findings, and a \emph{criticality-based} strategy, which ranks functions by their number of High- and Catastrophic-severity findings. Starting from the full assessment, functions were progressively removed in ascending order of rank — that is, least prevalent (or least critical) functions first — preserving higher-ranked functions in the reduced assessment.
Assessment burden was measured by the number of standardized CIS Controls v8 questionnaire items associated with the retained functions. Because each validated finding maps to one or more CIS Controls, and each control corresponds to one or more standardized questions, this mapping quantifies burden in terms of the questionnaire's structure rather than individual client responses. We focus specifically on this questionnaire component — the self-assessment portion of CIS Controls v8, as implemented in tools such as CISA's CSET \cite{cisa_cset} — rather than on technical evaluation activities (e.g., vulnerability scanning, configuration analysis, penetration testing), which require specialized tools and expertise beyond what most SMEs can perform independently. While SMEs can in principle complete the questionnaire without external expertise, the full 93-question instrument is often too extensive to complete unassisted; quantifying burden over this component allows a reduced function subset to inform a more tractable, potentially self-serve starting point for SME security assessment.

Thus, to evaluate what a reduction in assessment burden costs in terms of retained security value, we define three complementary coverage measures — finding coverage, critical-finding coverage, and probabilistic risk-pathway coverage — each capturing a different aspect of what a reduced assessment preserves relative to the full assessment.
First, we define \emph{finding coverage} as the proportion of validated assessment findings associated with the retained organizational security functions:
\begin{equation}
C_{\mathrm{find}}(S)
=
\frac{
N_{\mathrm{retained\ findings}}
}{
N_{\mathrm{total\ findings}}
},
\label{eq:finding_coverage}
\end{equation}
where $S$ denotes the set of organizational security functions retained in the reduced assessment.

Second, we define \emph{critical-finding coverage} as the proportion of High- and Catastrophic-severity findings associated with the retained functions:
\begin{equation}
C_{\mathrm{critical}}(S)
=
\frac{
N_{\mathrm{retained\ (High+Catastrophic)}}
}{
N_{\mathrm{total\ (High+Catastrophic)}}
}.
\label{eq:critical_coverage}
\end{equation}
Third, we define \emph{probabilistic risk-pathway coverage} as the proportion of the high-probability substantive pathway mass represented by the propagation model (Section~\ref{method:m1}) that is preserved by a reduced assessment. We use the complete exposure--attack--outcome pathway probabilities defined in Eq.~\ref{eq:path_probability}. Pathways containing an N/A attack state were excluded because they represent enabling conditions rather than substantive attack mechanisms. For each organizational security function, the remaining pathways were ranked in descending order of their model pathway probability. Because excluding N/A pathways can reduce the remaining probability mass below one, the non-N/A pathway probabilities were normalized within each function. The highest-probability pathways were then retained until they accounted for at least 80\% of the normalized non-N/A pathway probability. We denote this high-probability pathway set for function $F$ by $\mathcal{D}_F$. To evaluate sensitivity to this threshold, we repeated the analysis using 70\% and 90\% cumulative-probability thresholds. To account for differences in the prevalence of organizational security functions, each function was assigned the empirical weight
\begin{equation}
P(F)
=
\frac{
N_F
}{
N_{\mathrm{total\ findings}}
},
\label{eq:function_weight}
\end{equation}
where $N_F$ is the number of validated findings associated with function $F$.
Accordingly, the global weight of pathway $\pi$ associated with function $F$ is
\begin{equation}
W(F,\pi)
=
P(F)\,P(\pi\mid F),
\label{eq:pathway_weight}
\end{equation}
where $P(\pi\mid F)$ is the model pathway probability obtained from
Eq.~\ref{eq:path_probability}. For a reduced assessment retaining the set of organizational security functions $S$, probabilistic risk-pathway coverage is defined as:
\begin{equation}
C_{\mathrm{path}}(S)
=
\frac{
\displaystyle
\sum_{F\in S}
\sum_{\pi\in\mathcal{D}_F}
W(F,\pi)
}{
\displaystyle
\sum_{F\in\mathcal{F}}
\sum_{\pi\in\mathcal{D}_F}
W(F,\pi)
},
\label{eq:pathway_coverage}
\end{equation}
where $\mathcal{F}$ denotes the complete set of organizational security functions. Thus, Eq.~\ref{eq:pathway_coverage} quantifies the proportion of high-probability substantive pathway mass retained by a reduced assessment. Assessment-question reduction was then evaluated against the three coverage measures in Eqs.~\ref{eq:finding_coverage}, \ref{eq:critical_coverage}, and \ref{eq:pathway_coverage} under both prioritization strategies, characterizing the trade-off between assessment burden and retained security coverage and its sensitivity to the function-prioritization criterion.

\section{Results}
\label{section:results}
This section presents our empirical results, organized by research question: prevalence and severity of organizational security functions (RQ1), the empirical Risk Manifestation Model (RQ2), propagated risk-manifestation pathways and their robustness (RQ3), and evidence-based assessment reduction (RQ4).

\subsection{Organizational Security Functions (RQ1)}\label{rq1}
Table~\ref{tab:rq1_summary} summarizes the prevalence, organizational distribution, and severity of the identified organizational security functions. Four functions — Access Control \& Authentication, Configuration Management, Vulnerability Management, and Governance \& Organizational Capacity — accounted for 76.6\% of the 281 findings and appeared across 16--21 of the 22 SMEs. Access Control \& Authentication had the highest within-organization finding frequency (median = 3; IQR = 2--6), while Data Protection \& Resilience and Infrastructure \& Network Security had the highest proportion of critical findings (66.7\% and 58.3\%, respectively). Below, we provide detailed characterizations of each security function.

\begin{table}[h]
\centering
\caption{Distribution of validated security findings across organizational security functions. Critical findings comprise High and Catastrophic severity findings.}
\label{tab:rq1_summary}
\small
\setlength{\tabcolsep}{2.0pt}
\begin{tabular}{p{3.5cm}cccc}
\toprule
\textbf{Security Function} &
\textbf{Findings} &
\textbf{Orgs} &
\textbf{Med. (IQR)} &
\textbf{Critical} \\
\midrule
Access Control \& Authentication &
89 (31.7) &
21/22 &
3 (2--6) &
47 (52.8) \\

Configuration Management &
44 (15.7) &
16/22 &
2 (2--4) &
21 (47.7) \\

Vulnerability Management &
44 (15.7) &
16/22 &
2 (2--4) &
23 (52.3) \\

Governance \& Organizational Capacity &
38 (13.5) &
18/22 &
2 (2--2) &
19 (50.0) \\

Infrastructure \& Network Security &
24 (8.5) &
12/22 &
2 (1--2) &
14 (58.3) \\

Data Protection \& Resilience &
18 (6.4) &
12/22 &
1 (1--2) &
12 (66.7) \\

Monitoring \& Detection &
14 (4.9) &
11/22 &
1 (1--1) &
4 (28.6) \\

Asset Management &
10 (3.6) &
9/22 &
1 (1--1) &
0 (0.0) \\
\bottomrule
\end{tabular}
\end{table}

\par\noindent\textbf{(I) Access Control \& Authentication: }This function was the most prevalent function, accounting for 31.7\% of findings and affecting 21 of 22 SMEs. Common findings included weak credential policies, inconsistent multi-factor authentication, shared or default administrative accounts, exposed or reused credentials, and inadequate account lifecycle management. Deficiencies in privilege provisioning, review, and revocation also resulted in dormant accounts, excessive permissions, and poorly managed service accounts. Together, these findings point to identity-governance weaknesses that create persistent opportunities for unauthorized access and privilege escalation.

\par\noindent\textbf{(II) Configuration Management: }This function was characterized by insecure defaults, disabled security controls, outdated protocols, and misconfigured services. These findings indicate inconsistent processes for establishing and maintaining secure configurations, allowing insecure settings and unnecessary services to persist.

\par\noindent\textbf{(III) Vulnerability Management: }This function included outdated software, unsupported systems, legacy firmware, and applications with known CVEs. These findings reflect gaps in vulnerability identification, prioritization, patching, and lifecycle management rather than isolated software defects.

\par\noindent\textbf{(IV) Governance \& Organizational Capacity: }This function included missing security policies and incident-response procedures, limited awareness programs, unclear security responsibilities, and weak third-party governance. These findings suggest limited organizational oversight and capacity to manage security consistently across operational areas.

\par\noindent\textbf{(V) Infrastructure \& Network Security: }This function involved exposed services, insufficient segmentation, insecure remote access, vulnerable network devices, and limited protective controls. These weaknesses indicate deficiencies in network architecture and defense-in-depth that increase exposure to unauthorized access.

\par\noindent\textbf{(VI) Data Protection \& Resilience: }Despite its lower prevalence, this function had the highest proportion of critical findings (66.7\%). Findings included insecure data storage, inadequate encryption, unrestricted sharing, weak backups, and insufficient recovery planning — collectively indicating substantial potential consequences for data confidentiality and operational continuity.

\par\noindent\textbf{(VII) Monitoring \& Detection: }  This function was characterized by incomplete audit logging, limited centralized log management, and insufficient continuous monitoring across enterprise and cloud environments. These findings indicate that monitoring capabilities remain underdeveloped within many SMEs despite their importance for timely incident detection and response.

\par\noindent\textbf{(VIII) Asset Management: }This function  primarily involved incomplete asset inventories, limited visibility into hardware and software, and inadequate lifecycle-management processes. Although no critical findings occurred in this function, these weaknesses can constrain vulnerability management, configuration management, and broader security governance.

\noindent\textit{Overall, the results show that SME security weaknesses cluster around a small set of recurring organizational capabilities---particularly identity governance, secure configuration, vulnerability management, and governance---rather than being uniformly distributed across functions. Critically, prevalence and severity diverge: the most frequently observed functions are not necessarily the most severe, as less frequent functions such as data protection and network security can carry disproportionate severity.}

\begin{figure}[t]
\centering
\small
\begin{tikzpicture}[
    node distance=1.2cm,
    box/.style={
        draw,
        rectangle,
        rounded corners,
        align=center,
        minimum width=3.2cm,
        minimum height=0.4cm
    },
    relation/.style={
        draw,
        rectangle,
        rounded corners,
        align=center,
        minimum width=1.9cm,
        minimum height=0.45cm,
        inner sep=1pt,
        fill=white,
        font=\scriptsize
    },
    arrow/.style={->, thick}
]

\node[box] (finding) {
Assessment Findings\\
(Documented Weaknesses and Evidence)
};

\node[box] (func) [below of=finding] {
Organizational Security Functions\\
(Access Control, Governance, etc.)
};

\node[box] (exp) [below of=func] {
Exposure Conditions\\
(Asset Exposure, Network Exposure)
};

\node[box] (attack) [below of=exp] {
Attack Mechanisms\\
(Credential Compromise, Exploitation, etc.)
};

\node[box] (outcome) [below of=attack] {
Cybersecurity Outcomes\\
(Unauthorized Access, Data Compromise, etc.)
};

\draw[arrow] (finding) --
    node[midway, left=2mm, relation] {Coded into}
    (func);

\draw[arrow] (func) --
    node[midway, left=2mm, relation] {Associated with}
    (exp);

\draw[arrow] (exp) --
    node[midway, left=2mm, relation] {Associated with}
    (attack);

\draw[arrow] (attack) --
    node[midway, left=2mm, relation] {Associated with}
    (outcome);

\end{tikzpicture}
\caption{Conceptual representation of the empirical Risk Manifestation Model.}
\label{fig:risk_model}
\end{figure}
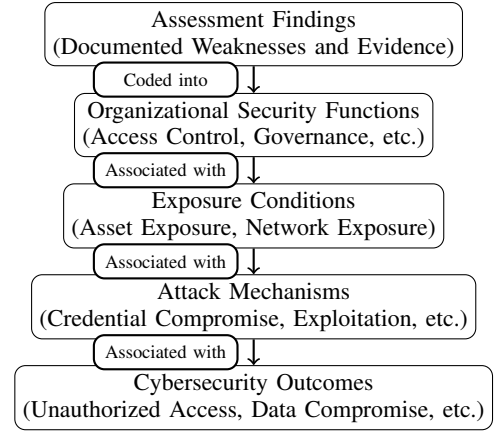

\subsection{Risk Manifestation Model \& Dependency Analysis (RQ2)}\label{rq2}
To characterize overall risk likelihood, we analyzed 254 of the 281 total findings, excluding cases without likelihood/impact annotations (primarily in CMMC Level 1 assessments and informational findings).
\par\noindent {\bf (I) Likelihood and Impact:} Most assessed risks were rated \textit{possible} (47.24\%) or \textit{likely} (24.02\%), with fewer rated \textit{unlikely} (18.11\%), \textit{rare} (9.06\%), or \textit{very likely} (1.57\%). Impact ratings were concentrated at the upper end: \textit{high} (49.21\%), \textit{moderate} (30.31\%), \textit{low} (12.60\%), \textit{catastrophic} (5.91\%), and \textit{insignificant} (1.97\%). Together, these ratings indicate that many assessed weaknesses combined plausible exploitation with substantial potential consequences. Because likelihood and impact characterize individual risks rather than how they manifest across organizational and technical layers, we next apply the Risk Manifestation Model (Section~\ref{method:m3}) to the organizational security functions identified in Section~\ref{rq1}. Figure~\ref{fig:risk_model} recaps the model's four-layer structure linking assessment findings to organizational security functions, exposure conditions, attack mechanisms, and cybersecurity outcomes.

\emph{Illustrative Example.} Consider a representative finding involving the absence of MFA on cloud-based platforms such as Google Drive and QuickBooks, rated \textit{likely} likelihood and \textit{high} impact given frequent phishing exposure and the sensitivity of accessible financial data. This weakness is associated with an asset exposure condition in which privileged accounts rely on single-factor authentication, supporting attack mechanisms such as credential compromise and human exploitation — for example, phishing that leads users to disclose credentials, enabling unauthorized access. These mechanisms are in turn associated with outcomes such as data exposure and reputational harm, illustrating the model's progression from organizational weakness to exposure, attack, and outcome.

\begin{figure*}[ht]
    \centering
    \includegraphics[width=\textwidth]{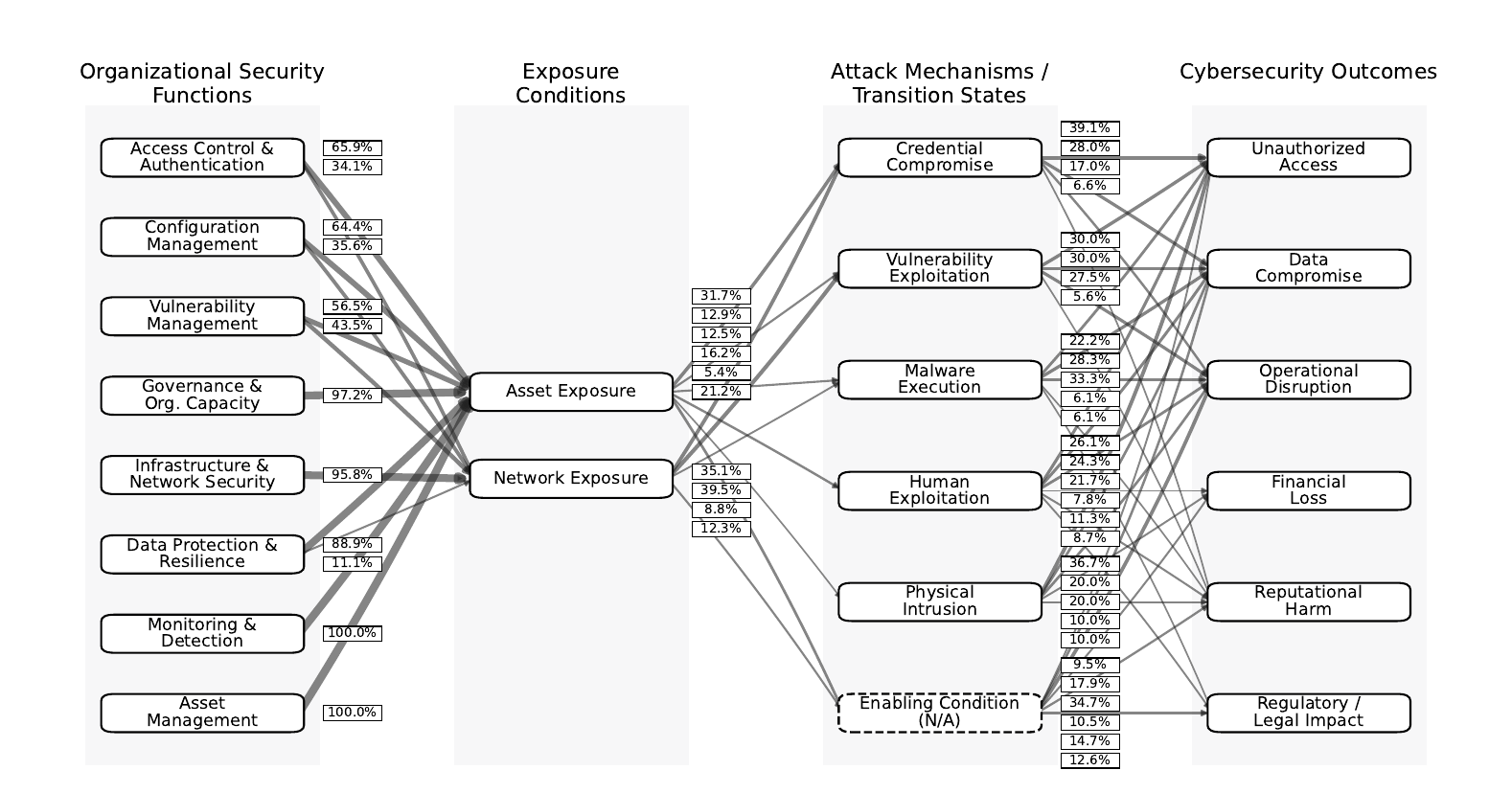}
    \caption{Empirical four-layer dependency graph linking organizational security functions, exposure conditions, attack mechanisms/transition states, and cybersecurity outcomes. Edge widths and labels represent conditional probabilities estimated from the coded assessment findings. Edges with conditional probabilities below 5\% are omitted for readability. The dashed node denotes an enabling condition rather than a substantive attack mechanism.}
\label{fig:dependency_graph}
\end{figure*}

\par\noindent {\bf (II) Empirical Dependency Analysis:} Using the coded assessment findings, we estimated the conditional relationships between organizational security functions, exposure conditions, attack mechanisms, and cybersecurity outcomes. Figure~\ref{fig:dependency_graph} summarizes the resulting empirical dependency graph. The graph shows how the eight organizational security functions identified in RQ1 correspond to two recurring exposure conditions---asset exposure and network exposure---which in turn are associated with five recurring attack mechanisms and six cybersecurity outcomes; a sixth, non-substantive attack-mechanism category labeled N/A in the dependency graph captures cases where an exposure condition does not correspond to a specific exploitation mechanism, and is excluded from the attack-mechanism-specific results discussed below. Edge widths and labels represent the estimated conditional probabilities between adjacent layers. Several clear patterns emerge from the graph. First, asset exposure dominates the propagation process for most organizational security functions, accounting for between 56.5\% and 100.0\% of the conditional probability mass for seven of the eight functions. Infrastructure \& Network Security is the primary exception, for which network exposure is the dominant downstream condition (95.8\%). Second, the dominant attack mechanism differs by exposure condition. Credential compromise is the most common attack mechanism following asset exposure (31.7\%), whereas vulnerability exploitation is the most common mechanism following network exposure (39.5\%), followed closely by credential compromise (35.1\%). Vulnerability exploitation is therefore substantially more prominent under network exposure than under asset exposure (12.9\%). At the outcome layer, unauthorized access receives the strongest incoming probability mass across multiple attack mechanisms, indicating that it is a recurring downstream cybersecurity outcome. Data compromise and operational disruption also receive substantial contributions from multiple attack mechanisms, whereas financial loss, reputational harm, and regulatory/legal impact occur less frequently and generally appear as secondary organizational consequences. These empirical dependencies provide the foundation for the probabilistic risk propagation analysis presented in RQ3. The complete conditional probability tables and full dependency graph are provided in Appendix~\ref{appendix:cpts}.

\noindent\textit{Overall, the observed dependencies indicate that organizational security weaknesses consistently correspond to a small set of recurring technical risk manifestations, motivating the probabilistic pathway analysis presented in RQ3.}



\begin{table*}[t]
\centering
\caption{Dominant risk-manifestation pathways with organization-level
jackknife uncertainty and leave-one-organization-out stability.}
\label{tab:dominant_pathways}
\small
\setlength{\tabcolsep}{1pt}

\begin{tabularx}{\textwidth}{
    >{\raggedright\arraybackslash}p{5.2cm}
    >{\raggedright\arraybackslash}p{9.6cm}
    >{\raggedright\arraybackslash}p{0.8cm}
    >{\raggedright\arraybackslash}p{1.5cm}
    >{\raggedright\arraybackslash}p{0.8cm}
}
\toprule
\textbf{Function} &
\textbf{Dominant Pathway} &
\textbf{Prob. (\%)} &
\textbf{95\% CI (\%)} &
\textbf{Stable (\%)} \\
\midrule

Access Control \& Authentication &
Asset Exposure $\rightarrow$ Credential Compromise
$\rightarrow$ Unauthorized Access &
8.17 & 3.88--12.38 & 100.0 \\

Configuration Management &
Asset Exposure $\rightarrow$ Credential Compromise
$\rightarrow$ Unauthorized Access &
7.98 & 4.27--11.66 & 100.0 \\

Vulnerability Management &
Asset Exposure $\rightarrow$ Credential Compromise
$\rightarrow$ Unauthorized Access &
7.00 & 4.14--9.91 & 100.0 \\

Governance \& Organizational Capacity &
Asset Exposure $\rightarrow$ Credential Compromise
$\rightarrow$ Unauthorized Access &
12.04 & 7.35--16.87 & 100.0 \\

Infrastructure \& Network Security &
Network Exposure $\rightarrow$ Credential Compromise
$\rightarrow$ Unauthorized Access &
13.15 & 8.68--17.70 & 100.0 \\

Data Protection \& Resilience &
Asset Exposure $\rightarrow$ Credential Compromise
$\rightarrow$ Unauthorized Access &
11.01 & 7.41--14.81 & 100.0 \\

Monitoring \& Detection &
Asset Exposure $\rightarrow$ Credential Compromise
$\rightarrow$ Unauthorized Access &
12.39 & 7.56--17.33 & 100.0 \\

Asset Management &
Asset Exposure $\rightarrow$ Credential Compromise
$\rightarrow$ Unauthorized Access &
12.39 & 7.56--17.33 & 100.0 \\

\bottomrule
\end{tabularx}
\end{table*}

\subsection{Probabilistic Risk Pathways \& Robustness Analysis (RQ3)}
\label{rq3}
Building on the empirical dependency graph presented in RQ2, we propagated conditional probabilities through the four-layer model to characterize the downstream risk pathways associated with each security function. This propagation yields both the attack mechanism distribution, $P(A|F)$, and the cybersecurity outcome distribution, $P(O|F)$, for each organizational security function. These propagated distributions are provided in Appendix~\ref{appendix:derived}, Tables~\ref{tab:function_attack} \&~\ref{tab:function_outcome}, while the main analysis focuses on the dominant end-to-end risk pathways. Table~\ref{tab:dominant_pathways} shows that seven of the eight organizational security functions share the same dominant pathway \textit{Asset Exposure $\rightarrow$ Credential Compromise $\rightarrow$ Unauthorized Access}. Infrastructure \& Network Security is the only exception, with its dominant pathway originating from \textit{Network Exposure}, consistent with its strong association with network exposure in RQ2. Dominant-pathway probabilities range from 7.00\% to 13.15\%, low in absolute terms given the large number of possible exposure–attack–outcome combinations, but consistently the highest-probability pathway for each function. Thus, the organizational security functions largely converge on a common risk-manifestation pathway, while differing in the amount of propagated probability mass assigned to that pathway.
\par\noindent{\bf {(I) Robustness Analysis:}} The LOO analysis yielded 176 function-level robustness comparisons. The dominant complete pathway remained unchanged in all 176 comparisons, corresponding to 100\% stability for every function shown in Table~\ref{tab:dominant_pathways}. The dominant cybersecurity outcome was also unchanged throughout, while the dominant marginal attack mechanism changed in only one comparison (99.43\% stability), for Infrastructure \& Network Security, without altering its dominant complete pathway. 
\par\noindent{\bf {(II) Uncertainty Estimation:}} To quantify sampling variability, the organization-level jackknife estimates yielded standard errors of 1.39--2.35 percentage points (mean: 2.02), with corresponding 95\% confidence intervals reported in Table~\ref{tab:dominant_pathways}. Despite variability in the probability estimates, the dominant pathway remained unchanged for every function, indicating robustness to the exclusion of any single organization.

\noindent\textit{Overall, probability propagation identifies a largely shared dominant risk-manifestation pathway across organizational security functions. Although its estimated probability varies under organization-level resampling, its identity remains stable across all LOO models.}

\begin{table}[t]
\centering
\caption{Assessment reduction under prevalence-based function prioritization.
Critical findings are High- or Catastrophic-severity findings.}
\label{tab:rq4}
\small
\setlength{\tabcolsep}{1pt}
\renewcommand{\arraystretch}{1.1}

\begin{tabular}{cccccc}
\toprule
\makecell{\textbf{Func.}\\\textbf{No.}} &
\makecell{\textbf{Asses.}\\\textbf{Qs}} &
\makecell{\textbf{Question}\\\textbf{Red. (\%)}} &
\makecell{\textbf{Findings}\\\textbf{Retained}} &
\makecell{\textbf{Critical}\\\textbf{Findings}} &
\makecell{\textbf{Pathway}\\\textbf{Cov. (\%)}} \\
\midrule

8 & 93 & 0.00 & 281 (100.00) & 140 (100.00) & 100.00 \\
7 & 79 & 15.05 & 271 (96.44) & 140 (100.00) & 96.55 \\

\textbf{6} &
\textbf{71} &
\textbf{23.66} &
\textbf{257 (91.46)} &
\textbf{136 (97.14)} &
\textbf{91.71} \\

\textbf{5} &
\textbf{51} &
\textbf{45.16} &
\textbf{239 (85.05)} &
\textbf{124 (88.57)} &
\textbf{85.49} \\

4 & 44 & 52.69 & 215 (76.51) & 110 (78.57) & 76.34 \\
3 & 26 & 72.04 & 177 (62.99) & 91 (65.00) & 63.21 \\
2 & 22 & 76.34 & 133 (47.33) & 68 (48.57) & 47.48 \\
1 & 13 & 86.02 & 89 (31.67) & 47 (33.57) & 31.77 \\

\bottomrule
\end{tabular}
\end{table}

\subsection{Evidence-Based Assessment Reduction (RQ4)}
\label{sec:rq4}
To examine whether the assessment can be reduced while preserving
security-relevant coverage, we evaluated nested function-level assessment subsets using the reduction procedure described in
Section~\ref{section:method}. Table~\ref{tab:rq4} reports the
prevalence-based hierarchy and the corresponding assessment-question
reduction, validated-finding coverage, critical-finding coverage, and
probabilistic risk-pathway coverage derived from RQ3. The analysis identifies two practically relevant operating points.
Retaining six functions reduces the assessment from 93 to 71 questions (23.66\%) while preserving 257 of 281 validated findings (91.46\%), 136 of 140 critical findings (97.14\%), and 91.71\% of risk-pathway coverage. This configuration therefore provides a \emph{security-oriented} reduction, removing nearly one quarter of the assessment while excluding only four critical findings. Retaining five functions provides a more aggressive \emph{efficiency-oriented} alternative, reducing the assessment to 51 questions (45.16\%) while retaining 239 of 281 validated findings (85.05\%), 124 of 140 critical findings (88.57\%), and 85.49\% of risk-pathway coverage. Moving from six to five functions removes 20 additional questions at a cost of 6.41, 8.57, and 6.22 percentage points in validated-finding, critical-finding, and risk-pathway coverage, respectively. Further reduction to four functions lowers the assessment to 44 questions (52.69\% reduction) while critical-finding and risk-pathway coverage fall to 78.57\% and 76.34\%, respectively — a disproportionate coverage loss relative to the modest additional burden reduction.

\begin{figure}[t]
    \centering
    \includegraphics[width=\columnwidth]
    {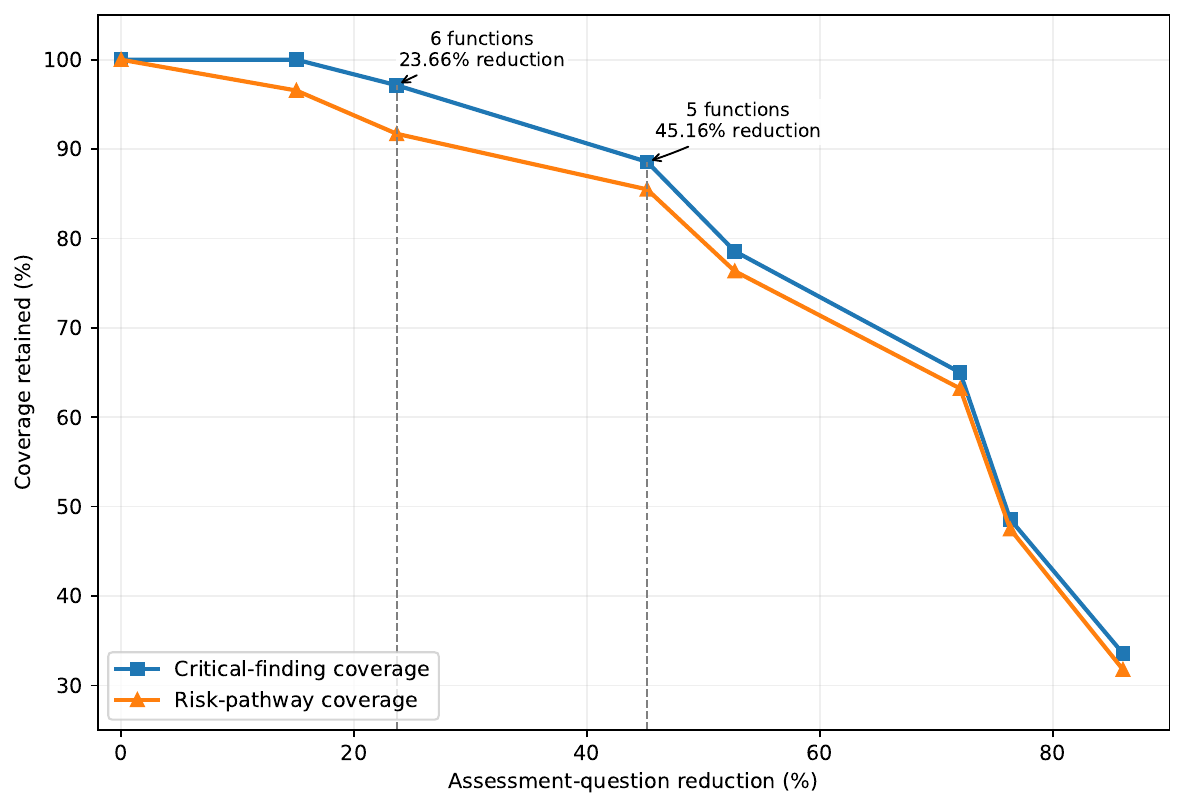}
    \caption{Assessment-question reduction versus retained security coverage.
    The marked six- and five-function configurations represent
    security-oriented and efficiency-oriented operating points, respectively.}
    \label{fig:rq4_tradeoff}
\end{figure}

\par\noindent\textbf{(I) Sensitivity to the prioritization criterion:}
We repeated the analysis by prioritizing functions according to
critical-finding prevalence rather than overall finding prevalence.
The rankings differed only in the ordering of Configuration Management and Vulnerability Management, which contained the same number of validated findings (44) but 21 and 23 critical findings, respectively. Consequently, the two strategies produced identical retained function sets and coverage results for all configurations retaining three or more functions, including the five- and six-function operating points; differences arose only at the
two-function configuration. Thus, the practically relevant
reduction--coverage trade-offs are robust to the prioritization criterion. Figure~\ref{fig:rq4_tradeoff} visualizes the reduction--coverage trade-off. Coverage declines gradually under moderate reduction but drops more noticeably between the six- and five-function configurations, illustrating the additional security-coverage cost associated with the larger reduction.

\noindent\textit{Overall, this empirical analysis indicates that meaningful reductions in assessment effort are possible while preserving most validated findings, critical findings, and probabilistic risk-pathway coverage. The six-function configuration provides a security-oriented reduction, whereas the five-function configuration offers a more efficiency-oriented alternative. Both operating points remain stable under critical-finding-based prioritization.}

\section{Discussion}
This study provides an empirical view of how recurring organizational security weaknesses in SMEs correspond to downstream technical risk. Across the four research questions, three broader implications emerge: diverse organizational weaknesses converge toward a comparatively small set of technical risk manifestations, network-facing weaknesses exhibit a distinct exposure profile, and this concentration enables substantial assessment reduction while preserving most observed security coverage.
\par\noindent\textbf{Convergence Toward Common Risk.} A central finding is the contrast between the diversity of organizational security weaknesses and the concentration of their downstream manifestations: of the eight recurring organizational security functions identified, seven shared the same dominant complete pathway, \emph{Asset Exposure $\rightarrow$ Credential Compromise $\rightarrow$ Unauthorized Access}, which remained unchanged across all 176 LOO comparisons. This convergence suggests that SME cybersecurity risk may be more concentrated at the technical manifestation layer than the diversity of organizational weaknesses initially implies. Prior SME research has emphasized resource limitations, governance challenges, staffing, and operational constraints \cite{chidukwani2022, rawindaran_2023, hoong2024navigating, wolf2021security}; our findings extend this perspective by empirically linking these organizational weaknesses to the exposure conditions, attack mechanisms, and cybersecurity outcomes observed during real-world assessments.
\par\noindent\textbf{Distinct Network-Facing Risk and Credential Convergence.} Infrastructure \& Network Security is the primary exception: its dominant pathway originates from \emph{network exposure} rather than \emph{asset exposure}, with vulnerability exploitation receiving substantially greater probability mass — consistent with externally reachable services, insecure configurations, and insufficient segmentation. Network-facing weaknesses should therefore not be treated as interchangeable with broader asset-related exposure when prioritizing assessment and mitigation. Credential compromise, meanwhile, remains prominent across otherwise different functions, marking it as a recurring mechanism through which diverse weaknesses manifest as downstream risk — an empirical point of convergence worth prioritizing, not a causal claim that identity controls eliminate risk from other weaknesses.
\par\noindent\textbf{Assessment Prioritization Under Resource Constraints.} The assessment-reduction analysis suggests that a substantial portion of the security coverage represented by the full assessment can be preserved with a smaller subset of organizational security functions — important for resource-constrained SMEs balancing assessment completeness with limited time, personnel, and expertise. Rather than a single preferred configuration, our results reveal two practically relevant operating points: a \emph{security-oriented} six-function configuration, preserving 97.14\% of critical findings while reducing assessment questions by 23.66\%, and a more \emph{efficiency-oriented} five-function configuration, preserving 88.57\% of critical findings while reducing questions by 45.16\%. Both operating points remained unchanged when functions were prioritized by critical-finding rather than overall prevalence, and pathway coverage varied by only 0.21–0.31 percentage points across 70\%, 80\%, and 90\% pathway-selection thresholds — indicating that the trade-off is not an artifact of the prioritization criterion or the primary analysis's threshold choice. The reduced configuration is therefore best viewed as an empirically supported assessment baseline for constrained settings, not a replacement for a comprehensive cybersecurity assessment. Because this baseline draws specifically on the self-administrable questionnaire component of CIS Controls v8, it may also offer a practical starting point for SMEs pursuing preliminary self-assessment before engaging external expertise.

\par\noindent\textbf{Limitations and Generalizability.}
Several limitations bound these conclusions. (1) This study is based on 22 SMEs assessed through a university cybersecurity clinic; although the organizations span multiple sectors, the results are not intended to be representative of the broader SME population and may not generalize to larger enterprises or different technological, geographic, or regulatory settings. Assessment scope also varied with client permissions — activities such as denial-of-service testing were not performed consistently, potentially underrepresenting some risk mechanisms. (2) The 281 findings are nested within organizations and should not be treated as statistically independent observations; function prevalence and pathway weights reflect the distribution of findings in this sample and may shift in other populations. (3) All findings originate from one clinic's assessment pipeline, coded primarily by a single researcher (with a second coder validating a 20\% subsample). Systematic biases in how students document findings could inflate the apparent dominance of particular pathways, and because organizational functions and downstream risk states are derived from the same source narratives, some cross-layer associations may partially reflect shared evidence rather than independent observations; findings with richer descriptions may also support multiple non-exclusive associations and thus contribute disproportionately to the probabilistic model. We mitigate these concerns through a structured codebook, independent coding and agreement analysis, evidence-supported edge construction, and LOO sensitivity analysis, though some annotation and reporting bias may remain. (4) The dependency graph captures empirical associations rather than causal relationships, and should not be interpreted as predicting future cybersecurity events. While LOO analysis showed high structural stability, confidence intervals around the estimated pathway probabilities remain relatively wide given the limited number of organizations; these results therefore demonstrate internal robustness but do not establish generalizability to unseen organizations, underscoring the need for prospective external validation on larger, more diverse datasets. 
\par\noindent\textbf{Future Work.} We plan to extend the dataset across additional organizations and contexts to evaluate whether the observed dependency structure generalizes, enabling a transition from descriptive characterization toward predictive modeling. We additionally plan to prospectively evaluate the assessment reduction on new clinic clients, investigate integration with CISA's CSET, and compare the resulting prioritization framework against established cyber-risk assessment approaches.

\section{Conclusion}
We presented a cross-layer empirical study linking organizational security weaknesses observed in SME cybersecurity assessments to exposure conditions, attack mechanisms, and cybersecurity outcomes. Using 281 validated findings from 22 real-world assessments, we identified eight recurring organizational security functions and constructed the Risk Manifestation Model, an empirical probabilistic structure derived from evidence-supported relationships in the assessment reports. The resulting analysis shows that diverse organizational weaknesses frequently converge on a small set of downstream technical risk manifestations — dominant pathways that remain stable under LOO analysis. Building on this structure, we show that substantially reducing assessment scope can preserve most of the full assessment's security coverage: a six-function configuration retains 97.14\% of critical findings while reducing assessment questions by 23.66\%, and a five-function configuration retains 88.57\% of critical findings while reducing questions by 45.16\%, with both operating points robust to the prioritization criterion and pathway-selection threshold used. Because the evaluated assessment component corresponds to the self-administrable portion of CIS Controls v8, these results may also inform more tractable, self-serve preliminary assessments for resource-constrained SMEs. Throughout, the modeled relationships represent empirical associations rather than causal or predictive claims; validating and extending this structure across a larger, more diverse population of organizations remains an important direction for future work. Overall, this study provides an empirical basis for connecting organizational cybersecurity conditions with technical risk and for prioritizing assessment effort in resource-constrained SMEs.


\section{Ethical Considerations}This study was reviewed by the Institutional Review Board (IRB) at XXX University (name anonymized for the review) and determined to be exempt as minimal-risk research. The study involves retrospective analysis of cybersecurity assessment reports generated through a university-led cybersecurity clinic. Any potentially identifying information, including organization names and contact details, was removed prior to analysis.
Results are reported using anonymized organizational identifiers; while findings are tracked at the organization level to support per-organization analyses such as LOO robustness checks, no organization's real-world identity is disclosed or inferable from the reported results.

\section{Open Science}
\label{appendix:open_science}
This study is based on real-world organizational cybersecurity assessment data. Due to the sensitive nature of these data and the risk of re-identification arising from a small and geographically concentrated sample, we are unable to publicly release client-level data or detailed finding descriptions that could compromise confidentiality. To support transparency and reproducibility, we will share the qualitative coding process, finding overviews, cycle 1 codes, cycle 2 codes, security functions, and the assessment-question-to-function mapping used in the analysis. These materials provide sufficient methodological transparency for readers to understand and evaluate the analytical process while protecting the confidentiality of participating organizations.




\bibliographystyle{IEEEtran}
\bibliography{ref}

\appendices


\onecolumn
\section{Illustrative Examples of Findings and Thematic Coding}
\label{sec:illustrative_examples}

Table~\ref{tab:appendix_example_finding} presents a representative
example of a security assessment finding from our dataset, illustrating
the structure and level of detail used in data collection.
\vspace{-0.5cm}
\begin{table*}[h]
\centering
\caption{Example security finding.}
\label{tab:appendix_example_finding}
\footnotesize
\renewcommand{\arraystretch}{1.1}
\setlength{\tabcolsep}{5pt}

\begin{tabularx}{\textwidth}{
    >{\raggedright\arraybackslash}p{3.2cm}
    >{\raggedright\arraybackslash}X
}
\toprule
\textbf{Field} & \textbf{Values} \\
\midrule

\textit{Identifier}
& ID F002 \\

\textit{Finding}
& Security Misconfiguration: Multifactor Authentication is Not Enabled \\

\textit{Finding Description}
& The business mainly uses Google Drive and QuickBooks for its business
functions and document storage. Given this reliance, using MFA to log in
to these services is essential for account and business security. This
is particularly important for accounts that access the most sensitive
information held on either platform. MFA should be enabled on all
accounts that access either platform, with particular emphasis on
privileged accounts. \\

\textit{Risk Rating}
& High \\

\textit{Likelihood}
& Likely -- The absence of MFA leaves accounts vulnerable to cyber
threats, including phishing attacks and unauthorized-access attempts.
The likelihood was rated as likely because compromise could occur if a
phishing attempt succeeds. If an attacker gains access to an account,
there would be no second authentication barrier preventing access.
Further, the client stated that it receives many spam emails and
phishing attempts each day, increasing the likelihood of compromise. \\

\textit{Impact}
& High -- A security breach resulting from the lack of MFA could
compromise business data, cause financial loss, and damage the
organization's reputation. Because QuickBooks is central to financial
operations, compromise could also cause severe financial repercussions
and operational disruption. \\

\textit{Remediation Effort Level}
& Low \\

\textit{Remediation Steps}
& Enable multifactor authentication on all business-sensitive cloud
platforms, including but not limited to Google Drive and QuickBooks. \\

\textit{CIS Control}
& 6.3: Require MFA for Externally Exposed Applications \\

\textit{Affected Assets}
& Cloud platforms -- Google Drive and QuickBooks \\

\textit{References}
& \url{https://www.nist.gov/itl/smallbusinesscyber/guidance-topic/multi-factor-authentication} \\

\textit{Supporting Evidence}
& Verbal interview with the client \\

\bottomrule
\end{tabularx}
\end{table*}

Table~\ref{tab:theme_mapping} provides examples of initial coding labels
to illustrate the level of detail and diversity of observations
underlying the thematic analysis.
\begin{table*}[h]
\centering
\caption{Mapping of initial coding labels to synthesized organizational security themes derived through iterative thematic analysis.}
\label{tab:theme_mapping}
\footnotesize
\renewcommand{\arraystretch}{1.1}
\setlength{\tabcolsep}{5pt}

\begin{tabularx}{\textwidth}{
    >{\raggedright\arraybackslash}p{0.70\textwidth}
    >{\raggedright\arraybackslash}p{0.25\textwidth}
}
\toprule
\textbf{Representative Initial Coding Labels}
& \textbf{Synthesized Security Functions} \\
\midrule

Identity verification (MFA, authentication); credential management
(passwords, default credentials); authorization (RBAC, least privilege);
account governance (account lifecycle, provisioning); administrative
access management; authentication policy enforcement; credential
exposure and misuse; session protection; device authentication
& Access Control \& Authentication \\

\addlinespace

Known software and firmware vulnerabilities (CVEs); end-of-life or
outdated software; patch and update management; continuous vulnerability
management processes; vulnerability assessment and remediation; weak or
obsolete security protocols and configurations; third-party application
and plugin vulnerabilities
& Vulnerability Management \\

\addlinespace

Secure configuration baselines; endpoint security configuration;
session and authentication configuration; removable-media controls;
network and protocol configuration; service and feature hardening;
application and web security configuration; malicious-code protection
configuration; configuration governance and policy
& Configuration Management \\

\addlinespace

Data governance and management; data classification and inventory;
secure data storage; encryption and data confidentiality; backup and
recovery; data access and sharing; sensitive-information exposure; data
lifecycle management
& Data Protection \& Resilience \\

\addlinespace

Incident response governance; security awareness and workforce
training; cybersecurity policies and governance; third-party and
external-service governance
& Governance \& Organizational Capacity \\

\addlinespace

Network boundary protection; network segmentation and architecture;
network service exposure; secure network protocols; wireless network
security; network traffic protection and resilience; external
connectivity management
& Infrastructure \& Network Security \\

\addlinespace

Audit logging; security event logging; log management and
centralization; security monitoring and analysis; incident detection
and investigation; incident response readiness
& Monitoring \& Detection \\

\addlinespace

No inventory of enterprise assets; outdated device inventory; unknown
devices; no software inventory; no allowed-software review; no
blacklist or whitelist; missing owner, department, hardware address,
approval status, or business purpose; no regular inventory review; no
support-status review; end-of-life software not identified; failure to
inventory physical-key distribution
& Asset Management \\
\bottomrule
\end{tabularx}
\end{table*}

\section{Conditional Probability Tables}
\label{appendix:cpts}

This appendix reports the empirical conditional probability tables
(CPTs) used to parameterize the probabilistic dependency graph presented
in Figure~\ref{fig:dependency_graph}. Each table corresponds to one layer
of the dependency graph: organizational security functions to exposure
conditions, $P(E\mid F)$; exposure conditions to attack mechanisms,
$P(A\mid E)$; and attack mechanisms to cybersecurity outcomes,
$P(O\mid A)$. Each cell reports the number of observed associations,
followed by the corresponding row-normalized percentage in parentheses.

\begin{table*}[h]
\centering
\caption{Conditional distribution of exposure conditions given
organizational security functions, $P(E\mid F)$.}
\label{tab:function_exposure}
\small
\renewcommand{\arraystretch}{1.2}
\setlength{\tabcolsep}{8pt}

\begin{tabular}{lcc}
\toprule
\textbf{Organizational Security Function}
& \textbf{Asset Exposure}
& \textbf{Network Exposure} \\
\midrule

Access Control \& Authentication
& 60 (65.9\%)
& 31 (34.1\%) \\

Configuration Management
& 29 (64.4\%)
& 16 (35.6\%) \\

Vulnerability Management
& 26 (56.5\%)
& 20 (43.5\%) \\

Governance \& Organizational Capacity
& 35 (97.2\%)
& 1 (2.8\%) \\

Infrastructure \& Network Security
& 1 (4.2\%)
& 23 (95.8\%) \\

Data Protection \& Resilience
& 16 (88.9\%)
& 2 (11.1\%) \\

Monitoring \& Detection
& 14 (100.0\%)
& 0 (0.0\%) \\

Asset Management
& 10 (100.0\%)
& 0 (0.0\%) \\

\bottomrule
\end{tabular}
\end{table*}

\begin{table*}[h]
\centering
\caption{Conditional distribution of attack mechanisms given exposure
conditions, $P(A\mid E)$.}
\label{tab:exposure_attack}
\small
\renewcommand{\arraystretch}{1.2}
\setlength{\tabcolsep}{6pt}

\begin{tabular}{lcccccc}
\toprule
\textbf{Exposure Condition}
& \makecell{\textbf{Credential}\\\textbf{Compromise}}
& \makecell{\textbf{Vulnerability}\\\textbf{Exploitation}}
& \makecell{\textbf{Malware}\\\textbf{Execution}}
& \makecell{\textbf{Human}\\\textbf{Exploitation}}
& \makecell{\textbf{Physical}\\\textbf{Intrusion}}
& \makecell{\textbf{N/A}\\\textbf{(Enabling Condition)}} \\
\midrule

Asset Exposure
& 76 (31.7\%)
& 31 (12.9\%)
& 30 (12.5\%)
& 39 (16.2\%)
& 13 (5.4\%)
& 51 (21.2\%) \\

Network Exposure
& 40 (35.1\%)
& 45 (39.5\%)
& 10 (8.8\%)
& 5 (4.4\%)
& 0 (0.0\%)
& 14 (12.3\%) \\

\bottomrule
\end{tabular}
\end{table*}

\begin{table*}[h]
\centering
\caption{Conditional distribution of cybersecurity outcomes given
attack mechanisms, $P(O\mid A)$.}
\label{tab:attack_outcome}
\small
\renewcommand{\arraystretch}{1.2}
\setlength{\tabcolsep}{6pt}

\begin{tabular}{lcccccc}
\toprule
\textbf{Attack Mechanism}
& \makecell{\textbf{Unauthorized}\\\textbf{Access}}
& \makecell{\textbf{Data}\\\textbf{Compromise}}
& \makecell{\textbf{Operational}\\\textbf{Disruption}}
& \makecell{\textbf{Financial}\\\textbf{Loss}}
& \makecell{\textbf{Reputational}\\\textbf{Harm}}
& \makecell{\textbf{Regulatory/}\\\textbf{Legal Impact}} \\
\midrule

Credential Compromise
& 106 (39.1\%)
& 76 (28.0\%)
& 46 (17.0\%)
& 13 (4.8\%)
& 18 (6.6\%)
& 12 (4.4\%) \\

Vulnerability Exploitation
& 48 (30.0\%)
& 48 (30.0\%)
& 44 (27.5\%)
& 5 (3.1\%)
& 9 (5.6\%)
& 6 (3.8\%) \\

Malware Execution
& 22 (22.2\%)
& 28 (28.3\%)
& 33 (33.3\%)
& 4 (4.0\%)
& 6 (6.1\%)
& 6 (6.1\%) \\

Human Exploitation
& 30 (26.1\%)
& 28 (24.3\%)
& 25 (21.7\%)
& 9 (7.8\%)
& 13 (11.3\%)
& 10 (8.7\%) \\

Physical Intrusion
& 11 (36.7\%)
& 6 (20.0\%)
& 6 (20.0\%)
& 3 (10.0\%)
& 3 (10.0\%)
& 1 (3.3\%) \\

N/A (Enabling Condition)
& 9 (9.5\%)
& 17 (17.9\%)
& 33 (34.7\%)
& 10 (10.5\%)
& 14 (14.7\%)
& 12 (12.6\%) \\

\bottomrule
\end{tabular}
\end{table*}

Together, Tables~\ref{tab:function_exposure}--\ref{tab:attack_outcome}
provide the complete empirical conditional probability tables used to
construct the dependency graph and perform the probabilistic risk
propagation analysis. These row-normalized conditional probabilities
parameterize the layered relationships between organizational security
functions, exposure conditions, attack mechanisms, and cybersecurity
outcomes.

\begin{figure*}[h]
    \centering
    \includegraphics[
        width=\textwidth
    ]{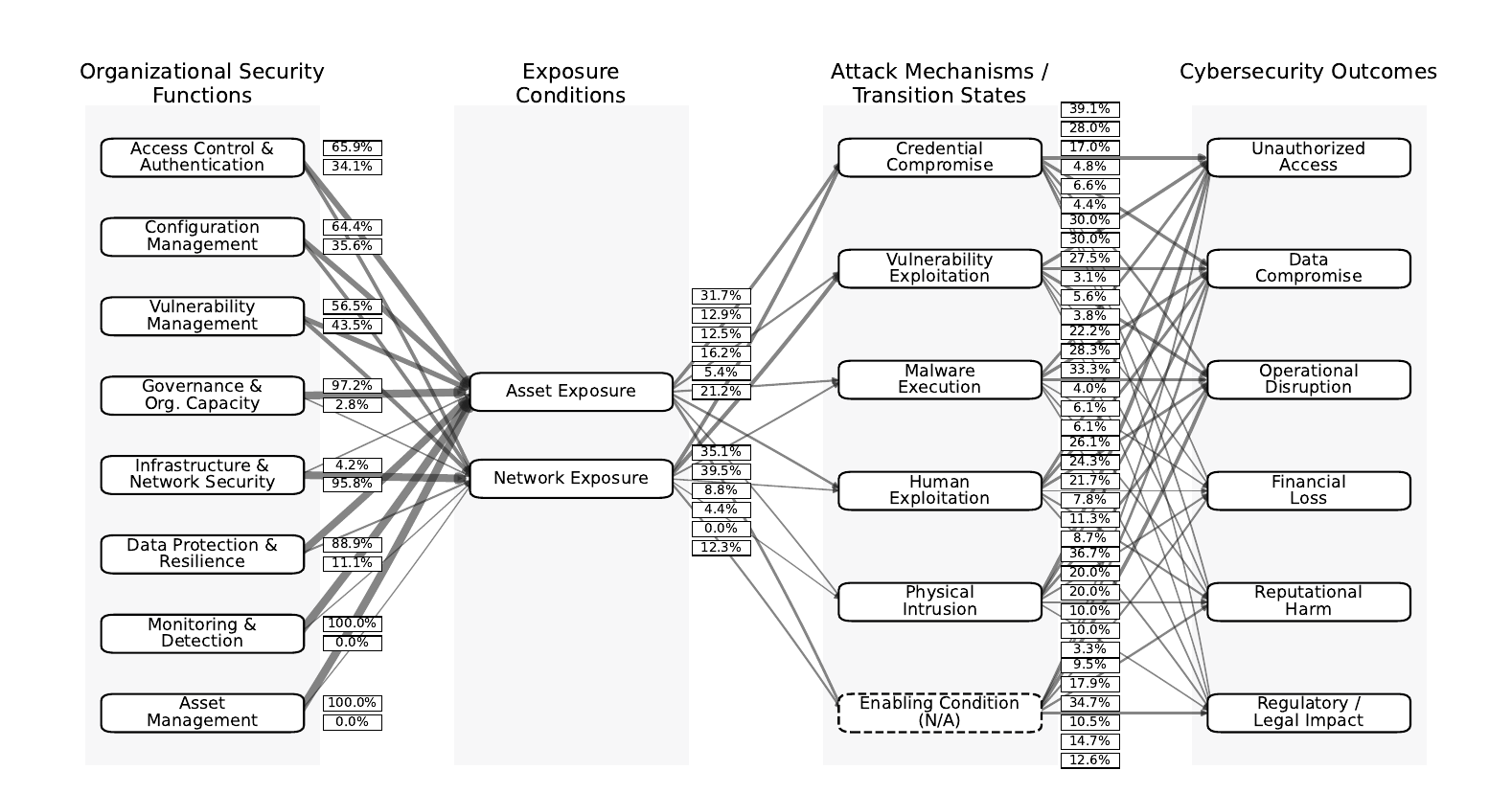}
    \caption{
    Full empirical four-layer dependency graph linking organizational
    security functions, exposure conditions, attack mechanisms, and
    cybersecurity outcomes. Edge widths and percentage labels represent
    conditional probabilities estimated from the coded assessment findings. The dashed
node denotes an enabling condition rather than a substantive attack mechanism.
    }
    \label{fig:dependency_graph_full}
\end{figure*}

Figure~\ref{fig:dependency_graph_full} presents the complete visualization of
the four-layer probabilistic dependency structure used in RQ2 and RQ3. The
graph links organizational security functions to exposure conditions, attack
mechanisms, and cybersecurity outcomes. Edge widths and percentage labels
represent the conditional probabilities estimated from the coded assessment
findings. 

\newpage
\section{Derived Probability Distributions}
\label{appendix:derived}

The probability propagation procedure combines the empirical conditional probability
tables reported in Appendix~\ref{appendix:cpts} to compute the propagated
attack-mechanism and cybersecurity-outcome distributions associated with
each organizational security function. Specifically,
\begin{equation}
P(A\mid F)
=
\sum_E P(A\mid E)P(E\mid F),
\label{eq:appendix_attack_inference}
\end{equation}
and
\begin{equation}
P(O\mid F)
=
\sum_E\sum_A P(O\mid A)P(A\mid E)P(E\mid F).
\label{eq:appendix_outcome_inference}
\end{equation}
Tables~\ref{tab:function_attack} and
\ref{tab:function_outcome} report these propagated probability
distributions. Unlike the empirical CPTs, these tables report derived
quantities obtained through probabilistic inference and are included for
completeness and reproducibility.

\begin{table*}[h]
\centering
\caption{Propagated attack-mechanism probability distribution for each
organizational security function, $P(A\mid F)$, computed using
Equation~\eqref{eq:appendix_attack_inference}. Values are percentages.}
\label{tab:function_attack}
\small
\renewcommand{\arraystretch}{1.2}
\setlength{\tabcolsep}{5pt}

\begin{tabular}{lcccccc}
\toprule
\textbf{Organizational Security Function}
& \makecell{\textbf{Credential}\\\textbf{Compromise}}
& \makecell{\textbf{Vulnerability}\\\textbf{Exploitation}}
& \makecell{\textbf{Malware}\\\textbf{Execution}}
& \makecell{\textbf{Human}\\\textbf{Exploitation}}
& \makecell{\textbf{Physical}\\\textbf{Intrusion}}
& \makecell{\textbf{N/A}\\\textbf{(Enabling Condition)}} \\
\midrule

Access Control \& Authentication
& 32.83 & 21.96 & 11.23 & 12.21 & 3.57 & 18.19 \\

Configuration Management
& 32.88 & 22.36 & 11.17 & 12.03 & 3.49 & 18.06 \\

Vulnerability Management
& 33.15 & 24.46 & 10.88 & 11.09 & 3.06 & 17.35 \\

Governance \& Organizational Capacity
& 31.76 & 13.65 & 12.40 & 15.92 & 5.27 & 21.00 \\

Infrastructure \& Network Security
& 34.95 & 38.37 & 8.93 & 4.88 & 0.23 & 12.65 \\

Data Protection \& Resilience
& 32.05 & 15.87 & 12.09 & 14.93 & 4.81 & 20.25 \\

Monitoring \& Detection
& 31.67 & 12.92 & 12.50 & 16.25 & 5.42 & 21.25 \\

Asset Management
& 31.67 & 12.92 & 12.50 & 16.25 & 5.42 & 21.25 \\

\bottomrule
\end{tabular}
\end{table*}

\begin{table*}[t]
\centering
\caption{Propagated cybersecurity-outcome probability distribution for
each organizational security function, $P(O\mid F)$, computed using
Equation~\eqref{eq:appendix_outcome_inference}. Values are percentages.}
\label{tab:function_outcome}
\small
\renewcommand{\arraystretch}{1.2}
\setlength{\tabcolsep}{5pt}

\begin{tabular}{lcccccc}
\toprule
\textbf{Organizational Security Function}
& \makecell{\textbf{Unauthorized}\\\textbf{Access}}
& \makecell{\textbf{Data}\\\textbf{Compromise}}
& \makecell{\textbf{Operational}\\\textbf{Disruption}}
& \makecell{\textbf{Financial}\\\textbf{Loss}}
& \makecell{\textbf{Reputational}\\\textbf{Harm}}
& \makecell{\textbf{Regulatory/}\\\textbf{Legal Impact}} \\
\midrule

Access Control \& Authentication
& 28.14 & 25.92 & 25.04 & 5.94 & 8.52 & 6.44 \\

Configuration Management
& 28.18 & 25.95 & 25.04 & 5.92 & 8.49 & 6.42 \\

Vulnerability Management
& 28.38 & 26.13 & 25.03 & 5.80 & 8.35 & 6.30 \\

Governance \& Organizational Capacity
& 27.35 & 25.20 & 25.09 & 6.43 & 9.05 & 6.88 \\

Infrastructure \& Network Security
& 29.72 & 27.33 & 24.96 & 4.97 & 7.46 & 5.56 \\

Data Protection \& Resilience
& 27.56 & 25.39 & 25.08 & 6.30 & 8.91 & 6.76 \\

Monitoring \& Detection
& 27.28 & 25.13 & 25.09 & 6.48 & 9.10 & 6.92 \\

Asset Management
& 27.28 & 25.13 & 25.09 & 6.48 & 9.10 & 6.92 \\

\bottomrule
\end{tabular}
\end{table*}

\end{document}